\documentclass[a4paper,11pt]{article}
\usepackage[usenames]{color}

\usepackage{amsfonts,amssymb}
\usepackage{theorem}
\usepackage[dvips]{graphicx}

\def\G{\Gamma}

\begin{document}
\par
\Large
\centerline{One-loop Self-energies in the Electroweak Model}
\centerline{with Nonlinearly Realized Gauge Group 
\footnote{\tt This work is supported in part by funds provided by the U.S. Department
of Energy (D.O.E.) under cooperative research agreement \#DE FG02-05ER41360}}
\normalsize
\rm 
\large
\vskip 1.1 truecm
\centerline{D.~Bettinelli$^c$\footnote{e-mail: {\tt daniele.bettinelli@mi.infn.it}},
R.~Ferrari$^{a,b}$\footnote{e-mail: {\tt ruggferr@mit.edu}},
A.~Quadri$^b$\footnote{e-mail: {\tt andrea.quadri@mi.infn.it}}}
\normalsize
\medskip
\begin{center}
$^a$
Center for Theoretical Physics\\
Laboratory for Nuclear Science\\
and Department of Physics\\
Massachusetts Institute of Technology\\
Cambridge, Massachusetts 02139
and\\
$^b$Dip. di Fisica, Universit\`a degli Studi di Milano\\
and INFN, Sez. di Milano\\
via Celoria 16, I-20133 Milano, Italy\\
$^c$
Physikalisches Institut, Albert-Ludwigs-Universit\"at\\  
Hermann-Herder-Strasse 3,  D-79104 Freiburg, Germany\\
(MIT-CTP-4015, IFUM-937-FT,  Freiburg PHENO-09/01, May  2009)
\end{center}
%
%
%
\centerline{\bf Abstract}

\rm
\begin{quotation}
We evaluate at one loop the selfenergies for  
the $W,Z$ mesons in the Electroweak
model where the gauge group is nonlinearly realized. In this model
the Higgs boson parameters are absent, while a second mass parameter appears
together with a scale for the radiative corrections.
We estimate these parameters in a simplified fit on leptons and gauge bosons data. 
We check physical unitarity and the absence of infrared divergences.
Landau gauge is used.
As a reference for future higher order computations the regularized 
$D$-dimensional amplitudes are provided. Eventually the limit $D 
\rightarrow 4$ is taken on physical
amplitudes.
\end{quotation}

\newpage
\section{Introduction}

A consistent strategy for the all orders 
subtraction of the divergences in nonlinearly
realized gauge theories has been recently developed
\cite{Bettinelli:2008ey}-\cite{Bettinelli:2007eu}
by extending some tools originally devised
for the nonlinear sigma model in the flat
connection formalism \cite{Ferrari:2005ii}.
The approach relies on the  local functional equation
for the 1-PI vertex functional \cite{Ferrari:2005ii}
(encoding the invariance of the group Haar
measure under local left transformations),
the weak power-counting theorem \cite{Ferrari:2005va} and the pure pole
subtraction of properly normalized 1-PI amplitudes
\cite{Bettinelli:2007zn}.
This scheme of subtraction fulfills all the relevant symmetries 
of the vertex functional. 
Physical unitarity is established as a consequence
of the validity of the Slavnov-Taylor identity
 \cite{Ferrari:2004pd}.

This strategy has been first applied to the 
nonlinearly realized $SU(2)$ massive Yang-Mills
theory \cite{Bettinelli:2007tq}.
The full set of one-loop counterterms and the 
self-energy have been obtained in \cite{Bettinelli:2007cy}.

The extension to the electroweak model based on the nonlinearly realized
$SU_L(2) \otimes U_Y(1)$ gauge group
introduces a number of additional non-trivial features
\cite{Bettinelli:2008ey,Bettinelli:2008qn}.
The direction of the Spontaneous Symmetry Breaking 
fixes the linear combination of the hypercharge and of
the third generator of the weak isospin giving rise to the
electric charge. 
Despite the fact that both the
hypercharge and the $SU_L(2)$ symmetry are non linearly
realized, the Ward identity for the electric charge has a linear
form on the vertex functional.

The anomalous couplings are forbidden by the $U(1)_Y$ invariance
together with the weak power-counting. However, two independent
mass parameters for the vector mesons are allowed. Thus the
ratio of the vector meson masses is not anymore given  by
the Weinberg angle.

As a first step toward a detailed analysis of the radiative
corrections of the nonlinearly realized electroweak theory,
we provide in this paper the vector meson self-energies
in the one-loop approximation.

The dependence of the self-energies on the second mass parameter 
is important in order to establish a comparison with the
linear realization of the electroweak group based on the
Higgs mechanism.

We provide a rough estimate both of the extra mass parameter
and of the scale of the radiative corrections. We fix some of the parameters
on measures taken at (almost) zero momentum transfer, while the
one-loop  corrections are confronted with measures at the
resonant value of the vector bosons energies.
The resulting values are challenging: the departure from the
Weinberg relation between the vector meson mass is very small
and the scale of the radiative corrections is of the order of hundred GeV. 
\par 
This is the aim of the work: to provide the amplitudes
in D dimensions for future high order computations and to provide a 
preliminary assessment of  the predictivity of the Electroweak Model
based on the nonlinearly realized gauge group 
including the one-loop self-energies corrections.
Electroweak physics is described 
with very reasonable parameters (the second mass parameter and the scale 
of the radiative corrections). 
\par
The cancellations among unphysical states 
required by Physical Unitarity can be easily traced out.
The physical amplitudes are shown to be free of infrared divergences.
It is remarkable also that they do not depend from  the Spontaneous Breakdown of Symmetry
parameter $v$. This fact has been discussed in Refs.  \cite{Bettinelli:2008ey},
\cite{Bettinelli:2008qn} and \cite{Bettinelli:2007cy}.

The computation is done in the symmetric formalism on the
$SU(2)_L$ flavor basis.
This choice simplifies greatly both the Feynman rules
and the actual computation; in fact symmetry arguments
turn out to be very useful in the calculation of the
invariant functions. The symmetric formalism puts emphasis on 
the fact that the entering parameters are not
renormalized (e.g. as in the on-shell renormalization procedure)
and are fixed at the end, by means of the comparison with the experimental
data.
Moreover the symmetric formalism
 makes the underlying symmetric structure encoded
by the local functional equation more transparent.

\section{Feynman rules}
\label{sec:feyn}
The classical action is written in order to establish the Feynman rules.
We omit all the external sources which are needed in order
to subtract the divergences at higher loops. We refer to the
previous publications \cite{Bettinelli:2008ey}, \cite{Bettinelli:2008qn} where the procedure is
described at length. The field content of the electroweak model based on
the nonlinearly realized $SU(2)_L\otimes U(1)$ gauge group
includes  the $SU(2)_L$
connection $A_\mu = A_{a\mu} \frac{\tau_a}{2}$
($\tau_a,~ a=1,2,3$ are the Pauli matrices), the 
$U(1)$ connection $B_\mu$, the fermionic left
doublets collectively denoted by $L$ and the right singlets, i.e.
\begin{eqnarray}
& L \in\Biggl\{ \left(
\begin{array}{r} l^u_{Lj}\\
l^d_{Lj}
\end{array} \right), \left(
\begin{array}{r}q^u_{Lj}\\
V_{jk}q^d_{Lk}
\end{array} \right), \quad j,k=1,2,3\Biggr\}, & \nonumber \\
& R \in\Biggl\{ \left(
\begin{array}{r}l^u_{Rj} \\
l^d_{Rj}
\end{array} \right), \left(
\begin{array}{r} q^u_{Rj}\\
q^d_{Rj}
\end{array} \right), \quad j = 1,2,3\Biggr\}. &
\label{sec.2.1}
\end{eqnarray} 
In the above equation the quark fields
$(q^u_j, j=1,2,3) = (u,c,t)$ and
$(q^d_j, j=1,2,3) = (d,s,b)$ are taken
to be the mass eigenstates in the tree-level
lagrangian; $V_{jk}$ is the CKM matrix.
Similarly we use for the leptons the notation
$(l^u_j, j=1,2,3) = (\nu_e,\nu_\mu,\nu_\tau)$ and
$(l^d_j, j=1,2,3) = (e,\mu,\tau)$.
The single left doublets are denoted by $L^l_j$,
$j=1,2,3$ for the leptons, $L^q_j$, $j=1,2,3$ for the 
quarks.
Color indices are not displayed.

One also introduces the $SU(2)$ matrix $\Omega$
\begin{eqnarray}
\Omega = \frac{1}{v} (\phi_0 + i \phi_a \tau_a) \, , ~~~
\Omega^\dagger \Omega = 1 \Rightarrow \phi_0^2 + \phi_a^2 = v^2 \, .
\label{sec.2.2}
\end{eqnarray}
The  mass scale $v$  gives $\phi$ the canonical dimension
at $D=4$. We fix the direction of 
Spontaneous Symmetry Breaking
by imposing the tree-level constraint
\begin{eqnarray}
\phi_0 = \sqrt{v^2 - \phi_a^2} \, . 
\label{sec.2.3}
\end{eqnarray}
The $SU(2)$ flat connection is defined by
\begin{eqnarray}
F_\mu = i \Omega \partial_\mu \Omega^\dagger \, .
\label{sec.2.4}
\end{eqnarray}
%

\subsection{Classical Action}

%
Discarding the neutrino mass terms, the 
classical action for the nonlinearly realized $SU(2) \otimes U(1)$ gauge group with 
two independent mass parameters
for the vector mesons can be written as follows,
where the dependence on $\Omega$ is explicitly shown:
\begin{eqnarray}&&
S =
\Lambda^{(D-4)} \int d^Dx\,\Biggl( \,2 \,Tr\, \biggl\{
- \frac{1}{4}  G_{\mu\nu} G^{\mu\nu} - \frac{1}{4}  F_{\mu\nu}
F^{\mu\nu}
\Biggr\}
\nonumber\\&&
+M^2 \,Tr\, \biggl\{\bigl (gA_{\mu}
- \frac{g'}{2} \Omega\tau_3 B_\mu \Omega^\dagger
- F_{\mu}\bigr)^2\biggr\}
\nonumber\\&& 
+M^2\frac{ \kappa }{2}\Bigl( Tr\bigl\{(g \Omega^\dagger A_\mu \Omega 
- g' B_\mu \frac{\tau_3}{2}
+ i \Omega^\dagger \partial_\mu \Omega) \tau_3\bigr\}\Bigr)^2
\nonumber\\&&
+\sum_L\biggl[
\bar L \bigr(i\not\!\partial +g\not\!\!A 
+\frac{g'}{2}Y_L\not\!B\bigl)L
+\sum_R\bar R \bigr(i\not\!\partial 
+ \frac{g'}{2} (Y_L+\tau_3) \not\!B\bigl)R
\biggr]
\nonumber \\&&
+\sum_j\biggl[
m_{l_j}~\bar R^l_j\frac{1-\tau_3}{2}\Omega^\dagger L^l_j
-
m_{q^u_j}~\bar R^q_j \frac{1+\tau_3}{2}\Omega^\dagger L^q_j
\nonumber\\&&
+
m_{q^d_k} V^\dagger_{kj} ~\bar R^q_k
\frac{1-\tau_3}{2}\Omega^\dagger 
 L^q_j +h.c.
\biggr]
\Biggr) \, .
\label{pre.1}
\end{eqnarray}
In $D$ dimensions the doublets $L$ and $R$ obey
\begin{eqnarray}
\gamma_{_D} L = - L \quad
\gamma_{_D} R =  R,
\label{pre.2}
\end{eqnarray}
being $\gamma_{_D}$ a gamma matrix that anticommutes with every other $\gamma^\mu$.

The non-Abelian field strength $G_{\mu\nu}$ is defined by
\begin{eqnarray}
G_{\mu\nu}= G_{a\mu\nu} \frac{\tau_a}{2} = 
(\partial_\mu A_{a\nu} - \partial_\nu A_{a\mu}
+ g \epsilon_{abc} A_{b\mu} A_{c\nu}) \frac{\tau_a}{2} \, ,
\label{n.ab.fs.}
\end{eqnarray}
while the Abelian field strength $F_{\mu\nu}$ is
\begin{eqnarray}
F_{\mu\nu} = \partial_\mu B_\nu - \partial_\nu B_\mu \, .
\label{ab.fs}
\end{eqnarray}
In the above equation the phenomenologically
successful  structure of the couplings
has been imposed by hand. However the same structure is required
by the Weak Power Counting requirement as discussed in 
Ref. \cite{Bettinelli:2008qn}.

\subsection{Gauge-fixing}

In order to set up the framework for the perturbative
quantization of the model,
the classical action in eq.(\ref{pre.1}) needs to be gauge-fixed.
The ghosts associated with the $SU(2)_L$ symmetry are denoted
by $c_a$. 
Their antighosts are denoted by $\bar c_a$, the Nakanishi-Lautrup
fields by $b_a$.
It is also useful to adopt the matrix notation
\begin{eqnarray}
c = c_a \frac{\tau_a}{2} \, , ~~~~ b = b_a \frac{\tau_a}{2} \, , ~~~~
\bar c =  \bar c_a \frac{\tau_a}{2} \, .
\label{sec.3.1}
\end{eqnarray}
The abelian ghost  is $c_0$, the abelian antighost $\bar c_0$
and the abelian Nakanishi-Lautrup field $b_0$.

For the sake of simplicity we deal here with the Landau gauge.
All external sources are  dropped out
since they are not relevant for the present work. The complete
set of external sources is provided in Ref. \cite{Bettinelli:2008qn}.
Then the gauge-fixing part of the classical action is
\begin{eqnarray}&&
S_{\rm GF}
\nonumber\\&&
=
\Lambda^{(D-4)}\int d^Dx \Biggl( b_0 \partial_\mu B^\mu 
-\bar c_0 \Box c_0 + 2 Tr~\Bigl\{  b\partial_\mu A^\mu 
-\bar c\partial^\mu D[A]_\mu c
\Bigr\} 
\Biggr) \nonumber \\
\label{pre.19}
\end{eqnarray}
%
\subsection{Bosons symmetric formalism}
%
The bilinear part of the boson sector is
\begin{eqnarray}&&
\frac{M^2}{2}\biggl[(gA_{a\mu}-g'B_\mu\delta_{3a}
-\frac{2}{v}\partial_\mu\phi_a)^2
+\kappa 
(G\, Z_\mu  -\frac{2}{v}\partial_\mu\phi_3)^2\biggr]
\nonumber\\&&
+ b_0 \partial_\mu B^\mu +b_a \partial_\mu A_a^\mu
\nonumber\\&&
=M^2  g^2|W_\mu^+-\frac{2}{vg}\partial_\mu\phi^+|^2  +b^+\partial^\mu W^-_\mu
+b^-\partial^\mu W^+_\mu
\nonumber\\&&
+\frac{M^2}{2} (G^2)(1+\kappa)\biggl[Z 
-\frac{2}{vG}\partial_\mu\phi_Z\biggr]^2
+ b_Z\partial^\mu Z_\mu +  b_A\partial^\mu A_\mu
\label{act.3.3.1}
\end{eqnarray}
We use the notations
\begin{eqnarray}&&
G= \sqrt{g^2+g'^2}, \qquad c= \frac{g}{G}, 
\qquad s= \frac{g'}{G}
\nonumber\\&&
M_W = gM, \qquad M_Z=\sqrt{ (1+\kappa)}G M
\label{se.2}
\end{eqnarray} 
and
\begin{eqnarray}\!\!\!\!\!\!\!\!
W^+ = \frac{1}{\sqrt 2}(A_1-iA_2),~
Z = \frac{1}{G}(gA_3 - g' B),~ 
A = \frac{1}{G}(g'A_3 + g  B)
\label{sym.1}
\end{eqnarray}
In the Landau gauge the propagator matrix for the 
bilinear form
\begin{eqnarray}-\frac{1}{ 4}F_{\mu\nu}F^{\mu\nu}+
\frac{m^2}{ 2}(V_\mu  -\frac{2}{v}\partial_\mu\phi_3)^2  +b \partial_\mu V^\mu
\label{sym.2}
\end{eqnarray}
 is given by
\begin{equation}
\left(
\begin{array}{llrr}
&V_\nu & b & \phi \\
V_\mu&\frac{-i}{p^2-m^2}(g_{\mu\nu}-\frac{p_\mu p_\nu}{p^2}) & 
-\frac{p_\mu}{p^2}&0\\
b&\frac{p_\nu}{p^2}& 0& -i\frac{v}{2p^2}\\
\phi&0
& -i\frac{v}{2p^2}&\frac{v^2}{4m^2}\frac{i}{p^2}
\end{array}
\right).
\label{fey.8}
\end{equation}
In the symmetric notation we have
\begin{eqnarray}
\langle (A_1A_1)_+\rangle& =& 
\langle (A_2A_2)_+\rangle= \langle (W^+ W^-)_+\rangle 
\nonumber\\
\langle (A_1A_2)_+\rangle & =& \langle (A_1A_3)_+\rangle=\langle (A_2A_3)_+\rangle
=0
\nonumber\\
\langle (A_3A_3)_+\rangle & =& \frac{1}{G^2}
\langle \biggl((gZ+g'A)(gZ+g'A)\biggr)_+\rangle
\nonumber\\ 
 & =&\frac{1}{G^2}
\biggl(g^2\Delta_{M_Z}+g^{'2}D \biggr)
\nonumber\\
\langle (A_3B)_+\rangle & =& \frac{1}{G^2}
\langle \biggl((gZ+g'A)(-g'Z+gA)\biggr)_+\rangle
\nonumber\\ 
 & =&\frac{gg'}{G^2}
\biggl(-\Delta_{M_Z}+D \biggr)
\nonumber\\
\langle (BB)_+\rangle & =& \frac{1}{G^2}
\langle \biggl((-g'Z+gA)(-g'Z+gA)\biggr)_+\rangle
\nonumber\\ 
 & =&\frac{1}{G^2}
\biggl(g^{'2}\Delta_{M_Z}+g^2 D \biggr),
\label{va.7}
\end{eqnarray}
where we used a short hand notation, e.g.
\begin{eqnarray}&&
\Delta_{M_Z}\longrightarrow \frac{-i}{p^2-M_Z^2}(g_{\mu\nu}-\frac{p_\mu p_\nu}{p^2})
\nonumber\\ &&
D  \longrightarrow \frac{-i}{p^2}(g_{\mu\nu}-\frac{p_\mu p_\nu}{p^2}).
\label{sym.7}
\end{eqnarray}
%
\subsection{Boson trilinear couplings}
For the one-loop calculation of the vector boson self-energies
one needs the usual Feynman rules and the trilinear
couplings generated by the two mass-invariants
in eq. (\ref{pre.1}). The first mass invariant generates the trilinear
couplings

\begin{eqnarray}&&
\frac{M^2}{2} \,
Tr\, \biggl\{2 \biggl (gA_{\mu}-\frac{g'}{2} \Omega B_\mu \,\tau_3\Omega^\dagger
- F_{\mu}\biggr)^2\biggr\}\biggr|_{\rm TRILINEAR}
\nonumber\\&&
=\frac{M^2}{2}\biggl\{
-4 \frac{G}{v^2} Z_{\mu}\epsilon_{3bc}\partial^\mu\phi_b\phi_c
-4 \frac{g}{v^2} \sum_{a=1,2}A_{a\mu}\epsilon_{abc}\partial^\mu\phi_b\phi_c
\nonumber\\&&
+ 4 \frac{gg'}{Gv}(-g'Z+gA)_\mu A_a^\mu\epsilon_{3ab}\phi_b
-8 \frac{g'}{Gv^2}(-g'Z+gA)_\mu \epsilon_{3bc}\partial^\mu\phi_b\phi_c
\biggr\}
\label{act.3.2}
\end{eqnarray}
The second mass invariant yields
\begin{eqnarray}&&
\frac{\kappa  M^2}{2}\biggl( Tr\biggl\{
 g \Omega^\dagger A_\mu \Omega\, \tau_3 - \frac{g'}{2} B_\mu
+i\Omega^\dagger\partial_\mu \Omega \,\tau_3\biggr\}\biggr)^2\biggl|_{\rm TRILINEAR}
\nonumber\\&&
=\frac{\kappa  M^2}{2}\biggl(
4\frac{1}{v^2}G\,Z_{\mu}\epsilon_{3bc}(\partial^\mu\phi_b)\phi_c
-4g\frac{1}{v}G\,Z^\mu A_{a\mu}\epsilon_{ab3}\phi_b
\nonumber\\&&
+\frac{8}{v^2}g A_{a\mu}\partial^\mu\phi_3\epsilon_{3ab}\phi_b
\biggr).
\label{act.3.4}
\end{eqnarray}
We put everything together
\begin{eqnarray}&&
S \biggl|_{\rm BOSON~TRILINEAR}
\nonumber\\&&
=\frac{M^2}{2}\Biggl\{
-4 \frac{g}{v^2} \sum_{a=1,2}A_{a\mu}\epsilon_{abc}\partial^\mu\phi_b\phi_c
+8\kappa\frac{g}{v^2} A_{a\mu}\partial^\mu\phi_3\epsilon_{3ab}\phi_b
\nonumber\\&&
+4 \frac{G}{v^2} \Bigl (\frac{-g^2+g'^2}{G^2}+\kappa\Bigr)Z_{\mu}
\epsilon_{3bc}\partial^\mu\phi_b\phi_c
-4 \frac{gG}{v}
(\frac{g'^2}{G^2}+\kappa)Z_\mu A_a^\mu\epsilon_{3ab}\phi_b
\nonumber\\&&
-8 \frac{gg'}{Gv^2}{A_\mu} \epsilon_{3bc}\partial^\mu\phi_b\phi_c
+ 4 \frac{g^2g'}{Gv}A_\mu A_a^\mu\epsilon_{3ab}\phi_b
\Biggr\}
\label{act.3.5}
\end{eqnarray}
%

%
\subsection{Fermion contribution}
\label{sec:fer}
The evaluation of the fermion loops requires a rule
on how to handle the $\gamma_5$ in dimensional regularization.
Our mechanism of removal of divergences is based on a regularization that
respects the local gauge invariance therefore $\gamma_5$ 
must anticommute with any $\gamma_\mu$. At one loop this is
possible, as it is well known, since there are no chiral anomalies.
For higher loop calculation any trace involving $\gamma_5$
must be considered as an independent amplitude up to the end
of the subtraction procedure. Eventually we evaluate the limit
at $D=4$ for physical amplitudes.
\par
The fermion contribution can be easily casted into
a global formula
\begin{eqnarray}&&
\Gamma_{\mu\nu}[ABST]\equiv 
-\langle 0|\biggl(\bar\psi(x)\gamma_\mu(A+B\gamma_5)\psi(x)
\bar\psi(0)\gamma_\nu(S+T\gamma_5)\psi(0)\biggr)_+|0\rangle
\nonumber\\&&
= - Tr\Biggl\{
\int \frac{dp}{(2\pi)^D}\int \frac{dq}{(2\pi)^D}
\gamma_\mu(A+B\gamma_5)\frac{\not\! p+\not\!q+m}{(q+p)^2-m^2}e^{ipx}
\nonumber\\&&
\gamma_\nu(S+T\gamma_5)\frac{\not\! q+M}{q^2-M^2}
 \Biggr\}
\label{fermion.5}
\end{eqnarray}
where $A,B,S,T$ are matrix elements corresponding to the flavor
and the color of the fermions with mass $m$ and $M$ and can be obtained from
the classical action (\ref{pre.1}). In particular the neutral sector
is
\begin{eqnarray}
G Z^\mu
\bar \psi\biggl[\biggl(\frac{\tau_3}{ 4}-s^2Q\biggr)\gamma_\mu 
-\frac{\tau_3}{ 4}\gamma_\mu\gamma_5\biggr] \psi
+e A^\mu\bar\psi Q\gamma_\mu\psi,  \qquad e\equiv \frac{gg'}{G}.
\label{fermion.22.1}
\end{eqnarray}
\par\noindent
One {then} gets {the transverse part of the contribution 
 of the fermions} (for the notations see Appendix \ref{app:A})
\begin{eqnarray}&&
\Gamma_{T}[ABST]
=4\frac{1}{D-1}
{\Huge Tr}\Biggl\{i(AS+BT)\frac{(2-D)}{2}\biggl(\Delta_m+\Delta_M\biggr)
\nonumber\\&&
+H(m,M)\biggl[mM(AS-BT)D+\frac{(2-D)}{2}(AS+BT)\biggl(-p^2+M^2+m^2\biggr)\biggr]
\nonumber\\&&
-\frac{1}{p^2}\frac{1}{2}(AS+BT)
(m^2-M^2)i(\Delta_M-\Delta_m)
\nonumber\\&&
-\frac{1}{p^2}H(m,M)\biggl[mM(AS-BT)p^2
\nonumber\\&&
+\frac{1}{2}(AS+BT)\biggl((m^2-M^2)^2-p^2(m^2+M^2)
\biggr)\biggr]
\Biggr\}
\label{fermion.14}
\end{eqnarray}
and {the longitudinal part of the contribution of the fermions}
\begin{eqnarray}&&
\Gamma_{L}[ABST]=\frac{4}{p^2}Tr
\Biggl\{
\frac{1}{2}(AS+BT)\Biggl(i(-m^2+M^2)\Delta_m
+i(m^2-M^2)\Delta_M
\Biggr)
\nonumber\\&&
+H(m,M)\biggl[mM(AS-BT)p^2
\nonumber\\&&
+\frac{1}{2}(AS+BT)\biggl(
M^4+m^4-2m^2M^2-p^2(m^2+M^2)
\biggr)\biggr]
 \Biggr\}.
\label{fermion.15}
\end{eqnarray}
%
\section{Self-energy amplitudes in $D$ dimensions}
\label{sec:se}
The presentation of the results (Landau gauge in D dimension) 
is as follows. First we report the result of the
calculation for the transverse and for the longitudinal
parts 
\begin{eqnarray}
\Sigma_{\mu\nu}=(g_{\mu\nu}-\frac{p_\mu p_\nu}{p^2})\Sigma_T 
+\frac{p_\mu p_\nu}{p^2}\Sigma_L.
\label{se.1}
\end{eqnarray} 
For each of them we report the contributions of  the single
graphs. Subsequently we evaluate the diagonal amplitudes $\Sigma_T$
on-shell, where we discuss the validity of physical unitarity and the
absence of any infrared singularity.
Finally the on-shell amplitudes are taken at $D=4$. 

We omit, for
sake of conciseness, 
to report the selfenergies at $D=4$ for generic momentum
(the procedure is straightforward).
We do not use on-shell renormalization: $M_W$ and $M_Z$ are
dummy parameters as well as $c$ and $s$.
The massless photon is a source of some infrared problems in the Landau gauge. 
{In Appendix \ref{app:A}} the analytical tools in order to handle these 
 difficulties {are provided (see eqs.(\ref{va.17.1.2}-\ref{va.17.1.8}))}. 
%

\subsection{The $D=4$ amplitudes}
%
The $D=4$ amplitudes are recovered as the finite
parts in the Laurent expansion of the generic 
dimensional regularized amplitudes, normalized by the factor
\begin{eqnarray}
\Lambda^{-(D-4)}.
\label{sm.1}
\end{eqnarray}
This point has been discussed at length in Refs. \cite{Ferrari:2005ii},
\cite{Bettinelli:2007zn}. 
In this procedure we encounter essentially the following cases.
\begin{eqnarray}
\Lambda^{-(D-4)}\Delta_m \Biggr|_{D\sim 4}
= \frac{m^2}{(4\pi)^2}\biggl(
\frac{2}{D-4} -1 +
\gamma + \ln\left[\frac{m^2}{4\pi\Lambda^2}\right]
\biggr),
\label{calc.4}
\end{eqnarray} 
\begin{eqnarray}&&
\Lambda^{-(D-4)}H(m,M)(p^2)\Biggr|_{D\sim 4}
=\frac{i}{(4\pi)^2}\Biggl\{\frac{2}{D-4}+\gamma
\nonumber\\&&
-\ln(4\pi)+ \int_0^1 dx 
\ln\biggl( \frac{m^2}{\Lambda^2}(1-x)+\frac{M^2}{\Lambda^2} x 
-\frac{p^2}{\Lambda^2}x(1-x)\biggr)\Biggr\}.
\label{calc.13.1}
\end{eqnarray} 
In Appendix \ref{app:A} we give the value of the last
integral in eq. (\ref{calc.13.1}).
\begin{eqnarray}&&
\Lambda^{-(D-4)}G(M)(p^2)\Biggr|_{D\sim 4}
=-\frac{i}{(4\pi)^2}
\Biggl[\frac{1}{p^2-M^2}\Bigl(\frac{2}{D-4}+\gamma -\ln 4\pi\Bigr)
\nonumber\\&& 
+\frac{1}{p^2-M^2}\Biggl(
\frac{M^2}{p^2}\ln\frac{[M^2-p^2]}{M^2}
+\ln \frac{[M^2-p^2]}{\Lambda^2}
\Biggr)\Biggr] 
.
\label{va.17.1.2.1}
\end{eqnarray} 

At $D\sim 4$ we use
\begin{eqnarray}
\frac{\Gamma(\frac{D}{2}-2)\Gamma(\frac{D}{2}-2)}{\Gamma(D-4)}
\simeq \frac{4}{D-4}+{\cal O}((D-4)^2) .
\label{ex.1}
\end{eqnarray} 
Then
\begin{eqnarray}&&
\Lambda^{4-D}\frac{\partial}{\partial M^2}G(M)\biggl|_{M=0}\sim
-\frac{i}{(4\pi)^2}[-p^2]^{-2}
\Bigl[1+(\gamma-1)(\frac{D}{2}-2)\Bigr]
\nonumber\\&&
\Bigl[1+(\frac{D}{2}-2)\log\Bigl(-\frac{p^2}{(4\pi)\Lambda^2}\Bigr)\Bigr]
\Bigl[\frac{4}{D-4}+{\cal O}((D-4)^2) \Bigr]
\nonumber\\&&
=-\frac{2i}{(4\pi)^2}[-p^2]^{-2}
\Biggl\{\frac{2}{D-4}-1+\gamma
+\log\Bigl(-\frac{p^2}{(4\pi)\Lambda^2}\Bigr)
\Biggr\}.
\label{ex.2}
\end{eqnarray} 

All other limit expressions can be reduced to eqs. (\ref{calc.4}-\ref{va.17.1.2.1})
\subsection{The counterterms}
The counterterms are given by the pole parts of the same Laurent 
expansion taken with a minus sign and finally multiplied
by the common factor $\Lambda^{(D-4)}$. 
In the expression (\ref{va.17.1.2.1}) the pole in $D-4$ dangerously multiplies
a nonlocal term. However we shall find that $G(M)$ always enter with
a factor $p^2-M^2$.
\section{WW selfenergy}
\label{sec:ww}
We first list the contributions to the transverse part
\subsection{Transverse WW-selfenergy}
%
The Goldstone bosons contribution to the transverse part of $\Sigma_{WW}$
\begin{eqnarray}&&
i\Sigma_{TWW}^{\rm GOLDSTONE}=
-i\frac{\Delta_{M_Z}}{4(D-1)}  M_W^2
\biggl[\frac{g^{'2}+\kappa  G^2}{G}\biggr]^2(\frac{1}{M_Z^2}+\frac{1}{p^2})
\nonumber\\&&
+\frac{G(0)}{D-1}  M_W^2\biggl[
\frac{gg'}{G}\biggr]^2
\frac{p^2}{4} 
+\frac{H(0,0)}{4(D-1)G^2 M_Z^2}
\Biggl\{-p^2 M_W^2\Bigl(g^{'2}+\kappa  G^2\Bigr)^2
\nonumber\\&&
+g^2  M_Z^2\biggl[
2(-3+2D)g'^2 M_W^2+G^2 p^2(1+\kappa) \biggr]
\Biggr\}
\nonumber\\&&
+\frac{H(0,M_Z)}{4G^2 }\biggl[g'^2+\kappa  G^2\biggr]^2 M_W^2
\Bigl(4+\frac{(M_Z^2- p^2)^2}{(D-1)M_Z^2 p^2}\Bigr)
\label{se.3}
\end{eqnarray} 
The Faddev-Popov contribution
\begin{eqnarray}
i\Sigma_{TWW}^{\rm FP}=
-\frac{g^2}{2(D-1)}p^2H(0,0).
\label{se.4}
\end{eqnarray} 
The vector boson tadpole contribution
\begin{eqnarray}
i\Sigma_{TWW}^{\rm TADPOLE}=i\frac{(D-1)^2}{D}g^2(\Delta_{M_W}+c^2\Delta_{M_Z}).
\label{se.5}
\end{eqnarray} 
The $\gamma W$ loop
\begin{eqnarray}&&
i\Sigma_{TWW}^{\gamma W}=
 -G(0)\frac{g^2 g'^2 p^6}{4 (D-1) G^2
    M_W^2} 
    -H(0,0)\frac{(D-2) g^2 g'^2  p^4}{(D-1) G^2
    M_W^2}  
\nonumber\\&&
    +G(M_W)\frac{g^2 g'^2 }{4 (D-1) G^2 M_W^2 p^2}
    \left(M_W^2-p^2\right)^2 \Bigl[M_W^4
\nonumber\\&&
+2 (2 D-3) p^2
    M_W^2+p^4\Bigr]  
\nonumber\\&&
    +H(0,M_W)\frac{g^2
    g'^2 }{(D-1) G^2 M_W^2 p^2}  \left(M_W^2+p^2\right) \Bigl[(D-2)
    M_W^4
\nonumber\\&&
+2 (3-2 D) p^2 M_W^2+(D-2) p^4\Bigr]
\nonumber\\&&
    -i \Delta_{M_W} \frac{g^2  g'^2 }
    {4 (D-1)D G^2 M_W^2 p^2}\Bigl(D (4 D-7)
    M_W^4+2 \Bigl[D (2 D-1)
\nonumber\\&&
-2\Bigr] p^2 M_W^2
+D (4 D-7) p^4\Bigr)
\label{se.6}
\end{eqnarray}
The $ZW$ loop
\begin{eqnarray}&& 
i\Sigma_{TWW}^{Z W}= 
\frac{g^4 H(0,0) p^6}{4 (D-1) G^2 M_W^2
    M_Z^2}
\nonumber\\&&
-\frac{g^4 H(0,M_W)}{4 (D-1)
    G^2 M_W^2 M_Z^2 p^2} \left(M_W^2-p^2\right)^2
   \Bigl[M_W^4+2 (2 D-3) p^2 M_W^2+p^4\Bigr]
\nonumber\\&&
-\frac{g^4 H(0,M_Z) }{4 (D-1) G^2 M_W^2 M_Z^2 p^2}
     \left(M_Z^2-p^2\right)^2 \Bigl[M_Z^4+2 (2 D-3) p^2
    M_Z^2+p^4\Bigr]
\nonumber\\&&
+\frac{i \Delta_{M_Z} g^4 }{4 (D-1) D
    G^2 M_Z^2 p^2}\Bigl\{
-D (4 D-7) p^4+\Bigl(D (4 D-7)
    M_W^2
\nonumber\\&&
-2\Bigl [D (2 D-1)-2 \Bigr] M_Z^2\Bigr) p^2+D\Bigl [M_W^4+(4
    D-7) M_Z^2 M_W^2
\nonumber\\&&
+(7-4 D) M_Z^4\Bigr]\Bigr\}
\nonumber\\&&
-\frac{i\Delta_{M_W}  g^4 }{4 (D-1) D G^2 M_W^2
    p^2}\Bigl\{ D (4 D-7)
    p^4+\Bigl(\left(4 D^2-2 D-4\right) M_W^2
\nonumber\\&&
+(7-4 D) D
    M_Z^2\Bigr) p^2+D\Bigl [ (4 D-7) M_W^4-(4 D-7) M_Z^2
    M_W^2
\nonumber\\&&
-M_Z^4\Bigr]\Bigr\}
\nonumber\\&&
+\frac{g^4 H(M_Z,M_W)}{4 (D-1) G^2 M_W^2 M_Z^2 p^2}
 ((M_W-M_Z)^2-p^2) ((M_W+M_Z)^2-p^2)
\nonumber\\&&
\Biggl\{M_W^4+2 (2
    D-3) \left(M_Z^2+p^2\right) M_W^2+M_Z^4+p^4+2 (2 D-3)
    M_Z^2 p^2\Biggr\}
\nonumber\\&&
\label{se.7}
\end{eqnarray}
The charged lepton contribution to transverse $\Sigma_{TWW}$
is obtained from eq. (\ref{fermion.14})
by using
\begin{eqnarray}
AS-BT=0, \qquad
AS+BT={\frac{g^2}{4}}.
\label{fermion.17.cl}
\end{eqnarray}
Therefore
\begin{eqnarray}&&
i\Sigma_{TWW}^{\rm LEPTONS}
=\frac{g^2}{2}
\frac{1}{D-1}\sum_{l=e,\mu,\tau}\Biggl\{(2-D)i\Delta_{M_l}+\frac{{M_l}^2}{p^2}
i\Delta_{M_l}
\nonumber\\&&
+H(0,M_l)(-p^2+{M_l}^2)\biggl[(2-D)-\frac{{M_l}^2}{p^2}\biggr]
\Biggr\}.
\label{fermion.18.T}
\end{eqnarray}
%
The quarks contribution to the transverse part of
$\Sigma_{WW}$ is given by eq. (\ref{fermion.14}) where
\begin{eqnarray}
AS-BT=0, \qquad
(AS+BT)_{ab}={\frac{g^2}{4}}3 V_{ab}V_{ab}^*
\label{fermion.17.cq}
\end{eqnarray}
being $V_{ab}$ the CKM matrix. Thus
\begin{eqnarray}&&
i\Sigma_{TWW}^{\rm QUARKS}=
3\frac{g^2}{2(D-1)}\sum_{ab}
V_{ab}V^*_{ab}\Biggl\{
i\Delta_{m_a}\biggl((2-D)+\frac{1}{p^2}
({m_a}^2-{M_b}^2)\biggr)
\nonumber\\&&
+i\Delta_{M_b}\biggl((2-D)-\frac{1}{p^2}
({m_a}^2-{M_b}^2)\biggr)
\nonumber\\&&
+H({m_a},{M_b})\biggl[(2-D)\bigl(-p^2+{M_b}^2+{m_a}^2\bigr)
\nonumber\\&&
-\frac{1}{p^2}\biggl(({m_a}^2-{M_b}^2)^2-p^2({m_a}^2+{M_b}^2)
\biggr)\biggr]
\Biggr\}
\label{fermion.20.T}
\end{eqnarray}
%
\subsection{Longitudinal WW selfenergy}
In a similar way we list the contributions for the longitudinal
parts of $\Sigma_{LWW}$.
The Goldstone bosons contribution to the longitudinal part of $\Sigma_{WW}$
\begin{eqnarray}&&
i\Sigma_{LWW}^{\rm GOLDSTONE}
= 
i \Delta_{M_Z} M_W^2\frac{\left(\kappa  G^2+g'^2\right)^2}{4G^2}
\left(\frac{1}{p^2}+\frac{1}{M_Z^2}\right)
\nonumber\\&&
   -G(0)M_W^2\frac{g^2  g'^2 p^2 }{4
    G^2}
    -H(0,M_Z)\frac{ \left(\kappa  G^2+g'^2\right)^2
     M_W^2}{4 G^2 M_Z^2
    p^2}\left(M_Z^2-p^2\right)^2
\nonumber\\&&
+\frac{1}{4} H(0,0) \left\{\frac{M_W^2 }{G^2 M_Z^2}
\left[2 g^2
    g'^2 M_Z^2+\left(\kappa  G^2+g'^2\right)^2
    p^2\right]
-\frac{g^2 \kappa ^2 p^2}{\kappa +1}\right\}.
\label{se.8.1.new}
\end{eqnarray} 
The Faddeev-Popov bosons contribution to the longitudinal part of $\Sigma_{WW}$
\begin{eqnarray}
i\Sigma_{LWW}^{\rm FP}=
-\frac{g^2}{2}p^2H(0,0).
\label{se.8.2.new}
\end{eqnarray} 
The vector boson tadpole contribution to longitudinal part
\begin{eqnarray}
i\Sigma_{LWW}^{\rm TADPOLE}=i\frac{(D-1)^2}{D}g^2(\Delta_{M_W}+c^2\Delta_{M_Z}).
\label{se.9}
\end{eqnarray} 
The $\gamma W$ loop contribution to longitudinal part
\begin{eqnarray}&&
i\Sigma_{LWW}^{\gamma W}=
 -G(W)\frac{g^2 g'^2 M_W^2 
    }{4 G^2 p^2}\left(M_W^2-p^2\right)^2
\nonumber\\&&
    -i \Delta_{M_W}\frac{ g^2g'^2
     }{2 G^2
    p^2}\left[\left(\frac{7}{2}-2 D\right) M_W^2+\left(2
    D-3+\frac{2}{D}\right) p^2\right]
\nonumber\\&&
+H(0,M_W)\frac{g^2   g'^2}{2 G^2 p^2}\left[-2 (D-2) M_W^4-p^2
    M_W^2+p^4\right]
\label{se.10}
\end{eqnarray} 
The $Z W$ loop contribution to longitudinal part
\begin{eqnarray}&&
i\Sigma_{LWW}^{Z W}=
\nonumber\\&&
 \frac{H(0,M_W) g^4}{4
    G^2 M_Z^2 p^2}\left(M_W^3-M_W p^2\right)^2 
+\frac{H(0,M_Z)g^4}{4 G^2 M_W^2 p^2} \left(M_Z^3-M_Z
    p^2\right)^2 
\nonumber\\&&
-\frac{i \Delta_{M_Z}
    g^4}{2 G^2 p^2}
    \Biggl\{\frac{ M_W^2}{2
    M_Z^2}(M_W^2-p^2) +\left(2 D-3+\frac{2}{D}\right) p^2
\nonumber\\&&
+\left(2
    D-\frac{7}{2}\right) (M_W^2-M_Z^2)\Biggr\} 
\nonumber\\&&
-\frac{i \Delta_{M_W} g^4}{2 G^2
    p^2}
    \Biggl\{\frac{M_Z^2}{2 M_W^2}(M_Z^2-p^2)+\left(2 D-3+\frac{2}{D}\right)
    p^2
\nonumber\\&&
+\left(2 D-\frac{7}{2}\right)
    \left(M_Z^2-M_W^2\right)\Biggr\} 
\nonumber\\&&
-\frac{H(M_Z,M_W)g^4}{4 G^2
    M_W^2 M_Z^2 p^2} \left(M_W^2-M_Z^2\right)^2
    \Biggl\{M_W^4+\left((4 D-6) M_Z^2-2 p^2\right)
    M_W^2
\nonumber\\&&
+\left(M_Z^2-p^2\right)^2 \Biggr\}
\label{se.11}
\end{eqnarray} 
The charged lepton contribution to longitudinal $\Sigma_{LWW}$
is obtained from eqs. (\ref{fermion.17.cl}) and  (\ref{fermion.15})
\begin{eqnarray}&&
i\Sigma_{LWW}^{\rm LEPTONS}=\frac{g^2}{2p^2}
\sum_{l=e,\mu,\tau}\Biggl\{-i M_l^2 \Delta_{M_l}
+H(0,M_l)(-p^2+{M_l}^2){M_l}^2
\Biggr\}.
\label{fermion.18.L}
\end{eqnarray}
The quark contribution to longitudinal $\Sigma_{LWW}$
is obtained from eqs. (\ref{fermion.17.cq}) and  (\ref{fermion.15})
\begin{eqnarray}&&
i\Sigma_{LWW}^{\rm QUARKS}= -
3\frac{g^2}{2p^2}\sum_{ab}
V_{ab}V^*_{ab}\Biggl\{
i\Delta_{m_a}
({m_a}^2-{M_b}^2)
\nonumber\\&&
-i\Delta_{M_b}
({m_a}^2-{M_b}^2)
\nonumber\\&&
-H({m_a},{M_b})\biggl[
({m_a}^2-{M_b}^2)^2-p^2({m_a}^2+{M_b}^2)\biggr]
\Biggr\}
\label{fermion.20.L}
\end{eqnarray}
%

%
\section{ZZ selfenergy}
\label{sec:ZZ}
We first list the contributions to the transverse part.
\subsection{The transverse ZZ selfenergy}
The Goldstone contribution to the transverse part of $ZZ$
selfenergy
\begin{eqnarray}&&
i\Sigma_{TZZ}^{\rm GOLDSTONE}
\nonumber\\&&
 =-i\Delta_{M_W}\frac{ \left(M_W^2+p^2\right) \left(\kappa
    G^2+g'^2\right)^2}{2 (D-1) G^2
    p^2}
\nonumber\\&&
+H(0,M_W)\frac{  \left(\kappa  G^2+g'^2\right)^2}{2 (D-1)
    G^2 p^2}\left[M_W^4+2 (2 D-3) p^2
    M_W^2+p^4\right]
\nonumber\\&&
+H(0,0)\frac{  p^2}{4 (D-1) G^2}
\left[g^4-2 \left(\kappa 
    G^2+g'^2\right) g^2-\left(\kappa 
    G^2+g'^2\right)^2\right]
\label{se.14}
\end{eqnarray} 
The Faddev-Popov contribution to the transverse part of $ZZ$
selfenergy
\begin{eqnarray}
i\Sigma_{TZZ}^{\rm FP}=
-\frac{g^4}{G^2}\frac{1}{2(D-1)}p^2 H(0,0).
\label{se.15}
\end{eqnarray} 
The vector boson tadpole contribution to transverse part
\begin{eqnarray}
i\Sigma_{TZZ}^{\rm TADPOLE}=
i\frac{g^4}{G^2}2\frac{(D-1)^2}{D}\Delta_{M_W}.
\label{se.16}
\end{eqnarray} 
The $W W$ loop contribution to transverse part of $ZZ$
selfenergy
\begin{eqnarray}&&
i\Sigma_{TZZ}^{W W}=
\nonumber\\&&
H(0,0)\frac{g^4  p^6}{4 (D-1) G^2 M_W^4}
\nonumber\\&&
-H(M_W,M_W)\frac{g^4
     \left(4 M_W^2-p^2\right) }{4 (D-1) G^2
    M_W^4}
    \left[4 (D-1) M_W^4+4 (2
    D-3) p^2 M_W^2+p^4\right]
\nonumber\\&&
-H(0,M_W)\frac{g^4  \left(M_W^2-p^2\right)^2
    }{2 (D-1)
    G^2 M_W^4 p^2}
    \left[M_W^4+2 (2 D-3) p^2 M_W^2+p^4\right]
\nonumber\\&&
-i \Delta_{M_W}\frac{ g^4 }{2 (D-1) D G^2
    M_W^2 p^2}
    \left[-D M_W^4+(5
    D-4) p^2 M_W^2+D (4 D-7) p^4\right]
\nonumber\\&&
\label{se.17}
\end{eqnarray} 
%
%
%
%
The neutrino contributions to transverse $\Sigma_{TZZ}$
is obtained from eq. (\ref{fermion.14})
by using
\begin{eqnarray}
AS-BT=0, \qquad
AS+BT={\frac{G^2}{8}}.
\label{fermion.17.cl.Z}
\end{eqnarray}
Therefore the neutrinos yield
\begin{eqnarray}
i\Sigma_{TZZ}^{\nu \nu}
=-3p^2\frac{G^2}{ 4}\frac{(2-D)}{D-1}H(0,0)
\label{fermion.23.1.Z}
\end{eqnarray}
For the charged leptons as well for the up- and down-quarks
the contribution to the selfenergy $\Sigma_{TZZ}$ has the same form
(\ref{fermion.14})
\begin{eqnarray}&&
i\Sigma_{TZZ}^{CHARGED~FERMIONS}
=\sum_j 4\frac{1}{D-1}\Biggl\{i(AS+BT)(2-D)\Delta_{m_j}
\nonumber\\&&
+H({m_j},{m_j})\biggl[2{m_j}^2\biggl((2-D)BT+AS\biggr)
-p^2\frac{(2-D)}{2}(AS+BT)
\biggr]
\Biggr\}
\label{fermion.18.T.Z}
\end{eqnarray}
where the sum is over the flavors. For the leptons
\begin{eqnarray}&&
AS-BT
=G^2s^2[-\frac{1}{ 2}+s^2]
\nonumber\\&&
AS+BT
=G^2[\frac{1}{8}-\frac{1}{ 2}s^2+s^4].
\label{fermion.24.Z}
\end{eqnarray}
For  up quarks
\begin{eqnarray}&&
AS-BT
=3G^2s^2[-\frac{1}{ 2}Q_u+s^2Q_u^2]
\nonumber\\&&
AS+BT
=3G^2[\frac{1}{8}-\frac{1}{ 2}s^2Q_u+s^4Q_u^2], \qquad
Q_u= \frac{2}{3}.
\label{fermion.25.Z}
\end{eqnarray}
For the down quarks
\begin{eqnarray}&&
AS-BT
=3G^2s^2[\frac{1}{ 2}Q_d+s^2Q_d^2]
\nonumber\\&&
AS+BT
=3G^2[\frac{1}{8}+\frac{1}{ 2}s^2Q_d+s^4Q_d^2],
\qquad
Q_d=- \frac{1}{3}.
\label{fermion.26.Z}
\end{eqnarray}
%

\subsection{The longitudinal ZZ selfenergy}
Now the longitudinal part of $ZZ$ selfenergy.
\par\noindent
The Goldstone contribution to the longitudinal part of $ZZ$ selfenergy
\begin{eqnarray}&&
i\Sigma_{LZZ}^{\rm GOLDSTONE}
=i \Delta_{M_W} \frac{
    \left(\kappa  G^2+g'^2\right)^2}{2
    G^2 p^2}\left(M_W^2+p^2\right) 
\nonumber\\&&
 + H(0,0)\frac{ \left(\kappa  G^2+g'^2\right)^2}{2
    G^2} p^2 
-H(0,M_W)\frac{ \left(\kappa 
    G^2+g'^2\right)^2}{2 G^2 p^2} \left(M_W^2-p^2\right)^2.
\label{se.14.1}
\end{eqnarray} 
The Faddev-Popov contribution to the longitudinal part of $ZZ$
selfenergy
\begin{eqnarray}
i\Sigma_{LZZ}^{\rm FP}=
-\frac{g^2c^2}{2}p^2H(0,0).
\label{se.15.1}
\end{eqnarray} 
The vector boson tadpole contribution to  the longitudinal part of $ZZ$
selfenergy
\begin{eqnarray}
i\Sigma_{LZZ}^{\rm TADPOLE}=
i\frac{g^4}{G^2}2\frac{(D-1)^2}{D}\Delta_{M_W}.
\label{se.16.1}
\end{eqnarray} 
The $W W$ loop contribution to the  longitudinal part of $ZZ$
selfenergy
\begin{eqnarray}&&
i\Sigma_{LZZ}^{W W}=
  H(0,M_W)  \frac{g^4}{2 G^2
    p^2}\left(M_W^2-p^2\right)^2
\nonumber\\&&
-i \Delta_{M_W} \frac{ g^4 }{2 G^2 p^2}\left\{M_W^2+\frac{1}{D}
\left[D (4 D-7)+4\right]    p^2\right\}.
\label{se.17.1}
\end{eqnarray} 
%
The fermion contribution to the longitudinal parts
of $\Sigma_{ZZ}$ is given by 
\begin{eqnarray}&&
i\Sigma_{LZZ}^{\rm FERMIONS}
=-8BT\sum_{j=leptons,quarks}m^2_j
H(m_j,m_j).
\label{fermion.24.1.L}
\end{eqnarray}
where $B,T$ is taken from eqs. (\ref{fermion.24.Z}), 
(\ref{fermion.25.Z}) and (\ref{fermion.26.Z}).
%
%
\section{$\gamma\gamma$ selfenergy}
%
We first list the contributions to the transverse part.
\subsection{The transverse $\gamma\gamma$ selfenergy}
%
\par\noindent
The Goldstone contribution to the transverse part of $\gamma\gamma$
selfenergy
\begin{eqnarray}&&
i\Sigma_{T\gamma\gamma}^{\rm GOLDSTONE}=
-i
    \Delta_{M_W} \frac{g^2  g'^2}{2 (D-1)
    G^2 p^2}\left(M_W^2+p^2\right)
+H(0,0) \frac{g^2 p^2 g'^2}{2 (D-1) G^2}
\nonumber\\&&
+ H(0,M_W)\frac{g^2  g'^2}{2 (D-1) G^2 p^2}
\left[M_W^4+2 (2 D-3) p^2
    M_W^2+p^4\right]
\label{se.18}
\end{eqnarray} 
The Faddev-Popov contribution to the transverse part of $\gamma\gamma$
selfenergy
\begin{eqnarray}
i\Sigma_{T\gamma\gamma}^{\rm FP}=
-\frac{e^2}{2(D-1)}p^2H(0,0).
\label{se.18.2.1}
\end{eqnarray} 
The vector boson tadpole contribution to transverse part
of $\gamma\gamma$ selfenergy
\begin{eqnarray}
i\Sigma_{T\gamma\gamma}^{\rm TADPOLE}=
i\frac{g^2g^{'2}}{G^2}2\frac{(D-1)^2}{D}\Delta_{M_W}
\label{se.19}
\end{eqnarray} 
The $W W$ loop contribution to transverse part of $\gamma\gamma$
selfenergy
%
\begin{eqnarray}&&
i\Sigma_{T\gamma\gamma}^{W W}=
H(0,0) \frac{g^2 g'^2  p^6}{4 (D-1) G^2 M_W^4}
-H(M_W,M_W)\frac{g^2 g'^2  }{4 (D-1) G^2 M_W^4}
\nonumber\\&&
\left(4    M_W^2-p^2\right) \left[4 (D-1)M_W^4+4 (2 D-3) p^2 M_W^2+p^4\right]
\nonumber\\&&
- H(0,M_W)\frac{g^2
    g'^2 }{2 (D-1) G^2M_W^4 p^2}\left(M_W^2-p^2\right)^2 \left[M_W^4+2
    (2 D-3) p^2M_W^2+p^4\right]
\nonumber\\&&
-i \Delta_{M_W}\frac{ g^2 g'^2 }{2 (D-1) D G^2M_W^2 p^2}\left[-DM_W^4+(5 D-4) p^2
   M_W^2+D (4 D-7) p^4\right]
\nonumber\\&&
\label{se.20}
\end{eqnarray} 
The electromagnetic interaction gives
\begin{eqnarray}
AS-BT=eQ, \qquad
AS+BT=eQ
\label{fermion.23.L}
\end{eqnarray}
then
\begin{eqnarray}&&
i\Sigma_{T\gamma\gamma}^{\rm FERMION}
=4\frac{e^2}{D-1}
\sum_{j=l,q,\rm color}Q^2
\Biggl\{i(2-D)\Delta_{m_j}
\nonumber\\&&
+\frac{-p^2(2-D)+4m_j^2}{2}H(m_j,m_j)
\Biggr\}.
\label{fermion.28.T}
\end{eqnarray}
For small $p^2$ one gets
\begin{eqnarray}&&
i\Sigma_{T\gamma\gamma}^{\rm FERMION}
= {\cal O}(p^2).
\label{fermion.29.T}
\end{eqnarray}
\subsection{
The longitudinal  $\gamma\gamma$ selfenergy.}
\par
The Goldstone contribution to the longitudinal part of $\gamma\gamma$
selfenergy
\begin{eqnarray}&&
i\Sigma_{L\gamma\gamma}^{\rm GOLDSTONE}
=
-\frac{1}{2p^2}M_W^2\biggl[\frac{gg^{'}}{G}\biggr]^2
\nonumber\\&&
\Biggl[ M_W^2(\frac{p^2}{M_W^2}-1)^2 H(0,M_W)
-\frac{p^4}{M_W^2}H(0,0)
-i\Delta_{M_W}(\frac{p^2}{M_W^2}+1)
\Biggr ]
\label{se.18.1}
\end{eqnarray} 
The Faddev-Popov contribution to the longitudinal part of $\gamma\gamma$
selfenergy
\begin{eqnarray}
i\Sigma_{L\gamma\gamma}^{\rm FP}=
-\frac{e^2}{2}p^2H(0,0).
\label{se.18.1.1}
\end{eqnarray} 
The vector boson tadpole contribution to the  longitudinal part
of $\gamma\gamma$ selfenergy
\begin{eqnarray}
i\Sigma_{L\gamma\gamma}^{\rm TADPOLE}=
i\frac{g^2g^{'2}}{G^2}2\frac{(D-1)^2}{D}\Delta_{M_W}
\label{se.19.1}
\end{eqnarray} 
The $W W$ loop contribution to the  longitudinal  part of $\gamma\gamma$
selfenergy
\begin{eqnarray}&&
i\Sigma_{L\gamma\gamma}^{W W}=
\frac{g^2g^{'2}}{G^2p^2}\Biggl\{
-i\Delta_{M_W}\biggl[
-\frac{p^2}{2}
+\frac{{M_W}^2}{2}+\frac{2p^2}{D}
+2Dp^2-3p^2\biggr]
\nonumber\\&&
+\frac{1}{2}H(0,{M_W})(p^2-{M_W}^2)^2\Biggr\}.
\label{se.20.1}
\end{eqnarray} 
For the longitudinal part we get
\begin{eqnarray}&&
i\Sigma_{L\gamma\gamma}^{\rm FERMION}
= 0.
\label{fermion.28.L}
\end{eqnarray}
It is remarkable that the sum of all the contributions (\ref{se.18.1})-(\ref{se.20.1})
amounts to zero  photon longitudinal self-energy. This is in agreement
with the Ward identity for QED derived in Ref. \cite{Bettinelli:2008ey}

%
\section{$Z\gamma $ selfenergy}
%
\label{sec:ZG}
We first list the contributions to the transverse part
\subsection{The transverse $Z\gamma $ selfenergy}
%
The Goldstone contribution to the transverse part of $Z\gamma$
selfenergy
\begin{eqnarray}&&
i\Sigma_{TZ\gamma}^{\rm GOLDSTONE}
\nonumber\\&&
=H(0,0) \frac{g'   g^3 p^2}{2 (D-1) G^2}
+i \Delta_{M_W} \frac{
    gg' \left(k G^2+g'^2\right)
     }{2 (D-1) G^2 p^2}\left(M_W^2+p^2\right)
\nonumber\\&&
-H(0,M_W)\frac{gg'
     \left(k G^2+g'^2\right) }{2 (D-1) G^2 p^2} \left[M_W^4+2 (2
    D-3) p^2 M_W^2+p^4\right]
\label{se.21}
\end{eqnarray}
The Faddev-Popov contribution to the transverse part of $Z\gamma$
selfenergy
\begin{eqnarray}
i\Sigma_{T Z\gamma}^{\rm FP}=
-\frac{g^3g^{'}}{G^2}\frac{1}{2(D-1)}p^2 H(0,0)
\label{se.24}
\end{eqnarray} 
The vector boson tadpole contribution to transverse part
of $Z\gamma$ selfenergy
\begin{eqnarray}
i\Sigma_{T Z\gamma}^{\rm TADPOLE}=
i\frac{g^3g^{'}}{G^2}2\frac{(D-1)^2}{D}\Delta_{M_W}
\label{se.23}
\end{eqnarray} 
The $W W$ loop contribution to transverse part of $Z\gamma$
selfenergy
%
{
\begin{eqnarray}&&
i\Sigma_{T Z\gamma}^{W W}=
 H(0,0)\frac{g^3 g'  p^6}{4 (D-1) G^2
    M_W^4}
\nonumber\\&&
    -H(M_W,M_W)\frac{g^3 g'  }{4 (D-1)
    G^2 M_W^4}\left(4 M_W^2-p^2\right)
\nonumber\\&&
    \Bigl[4 (D-1) M_W^4+4 (2 D-3) p^2 M_W^2+p^4\Bigr]
\nonumber\\&&
-H(0,M_W)\frac{g^3 g' 
    }{2 (D-1) G^2 M_W^4 p^2}\left(M_W^2-p^2\right)^2 \left[M_W^4+2 (2 D-3) p^2
    M_W^2+p^4\right]
\nonumber\\&&
-i\Delta_{M_W}\frac{ g^3 g' }{2 (D-1) D G^2 M_W^2 p^2}\left[-D M_W^4+(5 D-4) p^2 M_W^2+D (4D-7) p^4\right]
\nonumber\\&&
\label{se.22}
\end{eqnarray} 
}
The fermion contribution to  the transverse part of $\Sigma_{Z\gamma}$
is
\begin{eqnarray}&&
i\Sigma_{TZ\gamma}^{\rm FERMION}
=\sum_{j=leptons,quarks,color} 4\frac{(AS)_j}{D-1}\Biggl\{i(2-D)\Delta_{m_j}
\nonumber\\&&
+H({m_j},{m_j})
\biggl(p^2\frac{(D-2)}{2} +2{m_j}^2\biggr)
\Biggr\}
\label{fermion.24.1.pT}
\end{eqnarray}
where the constants $A,B,S,T$ are:
Neutrinos
\begin{eqnarray}
AS-BT=0, \qquad
AS+BT=0
\label{fermion.23.p}
\end{eqnarray}
Charged leptons
\begin{eqnarray}
AS-BT
=
AS+BT
=eG[\frac{1}{4}-s^2]
\label{fermion.24.p}
\end{eqnarray}
Up-Quarks
\begin{eqnarray}
AS-BT
=
AS+BT
=eQ_uG[\frac{1}{4}-s^2Q_u]
\qquad
Q_u= \frac{2}{3}
\label{fermion.25.p}
\end{eqnarray}
Down-Quarks
\begin{eqnarray}
AS-BT
=
AS+BT
=-eQ_dG[\frac{1}{4}+s^2Q_d]
\qquad
Q_d=- \frac{1}{3}
\label{fermion.26.p}
\end{eqnarray}
\subsection{The longitudinal $Z\gamma$ selfenergy}
\par\noindent
The Goldstone contribution to the longitudinal part 
of $Z\gamma$ selfenergy
\begin{eqnarray}&&
i\Sigma_{LZ\gamma}^{\rm GOLDSTONE}
=
\frac{1}{2}\frac{1}{p^2}\biggl[M_W^2\biggl(\frac{g^{'2}}{G}
+\kappa  G\,\biggr)\frac{gg^{'}}{G}\biggr]
\nonumber\\&&
\Biggl\{ M_W^2(\frac{p^2}{M_W^2}-1)^2 H(0,M_W)
-\frac{p^4}{M_W^2}H(0,0)
-i\Delta_{M_W}(\frac{p^2}{M_W^2}+1)
\Biggr\}
\label{se.21.1}
\end{eqnarray} 
The Faddev-Popov contribution to the longitudinal part of $Z\gamma$
selfenergy
\begin{eqnarray}
i\Sigma_{T Z\gamma}^{\rm FP}=
-\frac{g^3g^{'}}{G^2}\frac{1}{2}p^2 H(0,0)
\label{se.24.1}
\end{eqnarray} 
The vector boson tadpole contribution to the longitudinal
part of $Z\gamma$ selfenergy
\begin{eqnarray}
i\Sigma_{L Z\gamma}^{\rm TADPOLE}=
i\frac{g^3g^{'}}{G^2}2\frac{(D-1)^2}{D}\Delta_{M_W}
\label{se.23.1}
\end{eqnarray} 
The $W W$ loop contribution to the  longitudinal  
part of $\gamma Z$ selfenergy
\begin{eqnarray}&&
i\Sigma_{L Z\gamma}^{W W}=
\frac{g^3g^{'}}{G^2p^2}\Biggl\{
-i\Delta_{M_W}\biggl[
-\frac{p^2}{2}
+\frac{{M_W}^2}{2}+\frac{2p^2}{D}
+2Dp^2-3p^2\biggr]
\nonumber\\&&
+\frac{1}{2}H(0,{M_W})(p^2-{M_W}^2)^2\Biggr\}
\label{se.22.1}
\end{eqnarray} 
For the longitudinal one has
\begin{eqnarray}&&
i\Sigma_{LZ\gamma}^{\rm FERMION}
=0
\label{fermion.24.1.pL}
\end{eqnarray}
%

\section{Physical unitarity for diagonal elements}
\label{sec:uni}
It is simple to trace, at the one-loop level,
the contributions due to unphysical modes
(Faddeev-Popov ghosts, Goldstone bosons and scalar
parts of the vector mesons). These contributions
have to cancel when we evaluate the transverse part
of the selfenergies on-shell. Here we show that
this cancellation works in generic $D$ dimension
when the transverse part is taken on-shell.
\subsection{$\Sigma_{TWW}$:
the unphysical $H(0,0),H(0,M_Z)$ and $G(M_W),G(0)$}
We collect the terms in the self-energy in eqs. (\ref{se.3})-(\ref{se.7}) 
proportional to $H(0,0)$ and $H(0,M_Z)$, 
in order to check physical unitarity. They must vanish at $p^2=M_W^2$
since they are due to the presence of unphysical modes (Goldstone and
longitudinal part of the vector bosons).
We get
\begin{eqnarray}&&
H(0,0) \frac{g^2(p^2-M_W^2)}{4 (D-1) G^2 (k+1)
   M_W^4}\Biggl\{g^4p^2( p^2+M_W^2)
\nonumber\\&&
- 2g^2g'^2(1 + k)M_W^2
     \Bigl[(-3 + 2D)M_W^2 + 2(-2 + D)p^2\Bigr] 
\nonumber\\&&
- 2g'^4(1 + k)M_W^2
     \Bigl[ (-3 + 2D)M_W^2 + 2(-2 + D)p^2\Bigr] \Biggr\}
\label{sumWW.2}
\end{eqnarray} 
and
\begin{eqnarray}&&
-H(0,M_Z)\frac{(M_W^2-p^2)}{4 (D-1)
    G^4 (k+1) M_W^4 p^2}
\nonumber\\&&
\Bigl[G^4
    (k+1)^2 M_W^4+2 (2 D-3) g^2 G^2 (k+1) p^2
    M_W^2+g^4 p^4\Bigr]
\nonumber\\&&
\Bigl\{\left[(2 k+1) M_W^2-p^2\right]
    g^2+2 g'^2 (k+1) M_W^2\Bigr\}
.
\label{sumWW.4}
\end{eqnarray} 
Thus they are zero on-shell.
Similarly one can prove that on-shell $p^2=M_W^2$
the coefficients of $G(0)$ and of $G(M_W)$ are zero
{ in generic $D$ dimensions.}
\subsection{$\Sigma_{TZZ}$ terms proportional to $H(0,0)$ and $H(0,M_W)$}
We collect the terms in the self-energy in eqs. (\ref{se.14})-(\ref{se.17})
proportional to $H(0,0)$ and $H(0,M_W)$.
%
%
\begin{eqnarray}&&
\frac{H(0,0)}{D-1}\frac{p^2}{4M_W^4}\frac{g^4}{G^2}\Bigl\{
p^4-M_Z^4
\Bigr\}.
\label{sumZZ.2}
\end{eqnarray} 
and similarly
\begin{eqnarray}&&
\frac{g^4H(0,M_W) }{2 (D-1)G^2  M_W^4 p^2}  \left(p^2-M_Z^2\right)
\nonumber\\&&
\left(2M_W^2-M_Z^2-p^2\right)
     \left[M_W^4+2 (2 D-3) p^2
    M_W^2+p^4\right].
\label{sumZZ.3}
\end{eqnarray} 
Thus again physical unitarity is working for the self-mass of $Z$.

\subsection{$\Sigma_{T\gamma\gamma}$: the Limit $p^2=0$}
We collect the terms in the self-energy in eqs. (\ref{se.18})-(\ref{se.20})
proportional to $H(0,0)$ and then we put  at $p^2=0$
in order to check that the photon remains with zero mass. 
One verifies that
\begin{eqnarray}&&
\frac{H(0,0)p^2}{D-1}\frac{g^2g^{'2}}{G^2}\Biggl\{
1 -\frac{1}{2}
+\frac{p^4}{4M_W^4}
-\frac{1}{2}
\Biggr\}\Biggr|_{p^2=0}\quad = \quad 0.
\nonumber\\&&
\label{sumGG.2}
\end{eqnarray} 
For the terms involving $H(0,M)$ one needs the identity
\begin{eqnarray}
H(0,M)=
\frac{i\Delta_M}{M^2}\biggl[1-\frac{p^2}{M^2}(D-4)\biggl]
+{\cal O}(p^4).
\label{sumGG.2.6}
\end{eqnarray} 
It is then straightforward to verify that
\begin{eqnarray}
\lim_{p^2=0}
i\Sigma_{T\gamma\gamma}=0.
\label{sumGG.3}
\end{eqnarray} 
Thus the mass of the photon remains null.
\subsection{Unitarity for the $\Sigma_{TZ\gamma}$}
The unitarity properties of $\Sigma_{TZ\gamma}$
are strictly connected to the process where this graph contributes
(e.g. $Z\to l+\bar l$). Thus  more graphs are necessary in order to verify
physical unitarity. This subject is outside the scope of the
present work.
%
%
%
\section{$W$ and $Z$ selfmasses}
\label{sec:wm}
By using the procedure of extracting the finite
parts from the $D$-dimensional amplitudes described
in Sec. \ref{sec:se} we evaluate the selfmasses for
$W$  and $Z$ bosons. Since we have already thoroughly
examined the properties of the amplitudes in $D$ dimensions
at the onshell momenta, the selfmasses
can be evaluated by any computer algorithm. We do not
reproduce the result in the present paper.

\section{Parameters fit}
\label{sec:fit}
In this section we provide an estimate
of the parameters introduced in the model. The parameters $g,g',M$
can be fixed by experiments that are essentially at low momentum
transfer as for instance: $\alpha, G_\mu$
and the $\nu-e$
scattering that provides a precise value of $\sin \theta_W$.
Our calculation of the selfenergies can  be checked on the
physics of the vector bosons $W,Z$. The physical masses
are the imput for the determination of the extra
parameters of the model: $\kappa $ and $\Lambda$.

\par
For the processes at nearly zero momentum 
transfer
we can use Particle Data Group \cite{Amsler:2008zz} values as
\begin{eqnarray}&&
\alpha = 1/137.0599911(46)
\nonumber\\&&
G_\mu = 1.16637(1) \cdot 10^{-5} GeV
\label{fit.1}
\end{eqnarray}
and from $\nu-e$ scattering \cite{Vilain:1994qy}
\begin{eqnarray}
\sin^2\theta_W = 0.2324\pm 0.011 .
\label{fit.2}
\end{eqnarray}
We get:
\begin{eqnarray}&&
g = \frac{e}{s}= \frac{\sqrt{4\pi\alpha}}{s}=0.6281
\nonumber\\&&
g'=  \frac{e}{c}=0.3456
\nonumber\\&&
M= \sqrt{ \frac{1}{{4\sqrt 2}  G_F}} = 123.11 ~GeV.
\label{fit.3}
\end{eqnarray}
With these inputs we can evaluate the  values for the
other two parameters by imposing the conditions 
on the mass corrections 
\begin{eqnarray}&&
(gM)^2 +  \Delta M_W^2 = (80.428\pm 0.039)^2 ~ GeV^2
\nonumber\\&&
M^2 G^2(1+\kappa) + \Delta M_Z^2
= (91.1876\pm 0.0021)^2 ~ GeV^2
.
\label{fit.4}
\end{eqnarray}
One gets
\begin{eqnarray}&&
\kappa = 0.0085
\nonumber\\&&
\Lambda = 283~  GeV.
\label{fit.44}
\end{eqnarray}
{The widths of the vector mesons obtained from the imaginary parts
of the self-energies (all fermions are taken massless but the top with
$M_{top} = 174.2 ~ GeV$) are 
\begin{eqnarray}
&& \G_Z = 2.203 ~ GeV ~~~ (\mbox{exp.} ~~~   (2.4952 \pm 0.0023) ~ GeV ) \nonumber \\
&& \G_W = 1.818 ~ GeV ~~~ (\mbox{exp.} ~~~   (2.141 \pm 0.041) ~ GeV ) \, .
\label{fit.width}
\end{eqnarray}
}

These values are quite encouraging for the calculation of further radiative
corrections. However one should  consider only the order of
magnitude of these numbers. In fact they depend strongly from the
value of $\sin^2 \theta_W$. Only a fit including other sensitive
quantities will be able to reduce their variability.

\section{Comparison with the Standard Model}
%

The comparison with the linear theory is performed in the limit
$\kappa~=~0$, i.e. the tree-level Weinberg's relation
between the masses of the intermediate vector mesons
W and Z holds.
The Standard Model self-energy corrections are given by the same diagrams of the nonlinear model evaluated at $\kappa ~=~ 0$ plus the amplitudes involving
one internal Higgs line. 
The latter have been collected in \cite{Bardin:1999ak}.


We list below the amplitudes contributing to the transverse part of the W self-energy with an internal Higgs line. 
The results are valid in the Landau gauge and in the limit $D=4$.
%
%
%
%
%
The Higgs tadpole is 
\begin{eqnarray}
i \Sigma_{T WW}^{\rm HIGGS~TAD} = \frac{i\,g^2}{4}\, \Delta_{m_H} \,.
\label{neweq.2}
\end{eqnarray}
The Higgs-gauge bubble is
\begin{eqnarray}
&&\!\!\!\!\!\!\!\!\!
i \Sigma_{T WW}^{\rm HIGGS~GAUGE} = -\frac{g^2}{12}
\Big[i\Big(1+\frac{M_W^2}{p^2}-\frac{m_H^2}{p^2}\Big)\Delta_{M_W} -i \frac{M_W^2}{p^2}\,\Delta_{m_H}
-\frac{2i}{(4\pi)^2}\, M^2_W
\nonumber\\&&~~~~~~~~~~~~~~~~~~~~
+\Big(\big(m^2_H-M_W^2\big)^2\,\frac{1}{p^2}+
p^2+10M^2_W-2m^2_H\Big)H(M_W,m_H)\nonumber\\&&~~~~~~~~~~~~~~~~~~~~~~~
-\Big(1-\frac{m^2_H}{p^2}\Big)^2\,p^2\, H(0,m_H) \Big]\,.
\label{neweq.3}
\end{eqnarray}
The Higgs-Goldstone bubble is
\begin{eqnarray}
&&\!\!\!\!\!\!\!\!\!
i \Sigma_{T WW}^{\rm HIGGS~GOLDSTONE} = -\frac{g^2}{12}
\Big[i\Big(1+\frac{m_H^2}{p^2}\Big)\Delta_{m_H} 
-\frac{2i}{(4\pi)^2}\,\Big(m^2_H-\frac{p^2}{3}\Big)
\nonumber\\&&~~~~~~~~~~~~~~~~~~~~~~~~~~
+\Big(1-\frac{m^2_H}{p^2}\Big)^2\,p^2\, H(0,m_H) \Big]\,.
\label{neweq.4}
\end{eqnarray}
%
%
%
We list here the various contributions to the transverse part
of the $Z$ self-energy with an internal Higgs line.
%
%
%
The Higgs tadpole is
\begin{eqnarray}
i \Sigma_{T ZZ}^{\rm HIGGS~TAD} = \frac{i\,G^2}{4}\, \Delta_{m_H} \,.
\label{neweq.6}
\end{eqnarray}
The Higgs-gauge bubble is
\begin{eqnarray}
&&\!\!\!\!\!\!\!\!\!
i \Sigma_{T ZZ}^{\rm HIGGS~GAUGE} = -\frac{G^2}{12}
\Big[i\Big(1+\frac{M_Z^2}{p^2}-\frac{m_H^2}{p^2}\Big)\Delta_{M_Z} -i \frac{M_Z^2}{p^2}\,\Delta_{m_H}
-\frac{2i}{(4\pi)^2}\, M^2_Z
\nonumber\\&&~~~~~~~~~~~~~~~~~~~~
+\Big(\big(m^2_H-M_Z^2\big)^2\,\frac{1}{p^2}+
p^2+10M^2_Z-2m^2_H\Big)H(M_Z,m_H)\nonumber\\&&~~~~~~~~~~~~~~~~~~~~~~~
-\Big(1-\frac{m^2_H}{p^2}\Big)^2\,p^2\, H(0,m_H) \Big]\,.
\label{neweq.7}
\end{eqnarray}
The Higgs-Goldstone bubble is
\begin{eqnarray}
&&\!\!\!\!\!\!\!\!\!
i \Sigma_{T ZZ}^{\rm HIGGS~GOLDSTONE} = -\frac{G^2}{12}
\Big[i\Big(1+\frac{m_H^2}{p^2}\Big)\Delta_{m_H}
-\frac{2i}{(4\pi)^2}\,\Big(m^2_H-\frac{p^2}{3}\Big)
\nonumber\\&&~~~~~~~~~~~~~~~~~~~~~~~~~~
+\Big(1-\frac{m^2_H}{p^2}\Big)^2\,p^2\, H(0,m_H) \Big]\,.
\label{neweq.8}
\end{eqnarray}
These results can be used in order to estimate the numerical
impact of the Higgs corrections to the self-masses.
We choose as a reference value $m_H = 165 ~ GeV$ and
evaluate the corrections with the same input
parameters in eq.(\ref{fit.3}) and  $\Lambda = 283 ~ GeV$. 
The shifts in the self-masses are
\begin{eqnarray}
&& \Delta M_W^{\rm HIGGS} = 0.629 ~ GeV \,  , \nonumber \\
&& \Delta M_Z^{\rm HIGGS} = 0.531 ~ GeV \, .
\label{neweq.8.1}
\end{eqnarray}
These estimates are rather intriguing. $\Delta M_W^{\rm HIGGS}$ and
$\Delta M_Z^{\rm HIGGS}$ strongly depend on 
the value of $\Lambda$ (they vary by more than 20\% 
in the range from $\Lambda = 200 ~ GeV$ to
$\Lambda = 350 ~ GeV$).
Compensations of the Higgs contributions 
to electroweak observables may be triggered  
by a change in the scale $\Lambda$ of the radiative corrections.
A more refined fit to the electroweak precision
observables is required in order
to discriminate between the linear and the nonlinear 
theory.

\section{Conclusions}
%
The one loop evaluation of selfenergies for the vector mesons in the
Electroweak model based on nonlinearly realized gauge group
has been explicitly performed in D dimensions.
The finite amplitudes in $D=4$ has been achieved according  
to the procedure suggested by the local functional equation
associated to the local invariance of the path integral measure.
In practice this implies the minimal subtraction of poles in $D-4$ on properly
normalized amplitudes. Thus in the model the Higgs sector
is absent and the parameters are fixed by the classical
lagrangian (no free parameters for the counterterms and therefore
no on-shell renormalization). 
Two new parameters appear: a second mass term parameter and
a scale of radiative corrections. The Spontaneous Symmetry  Breaking
parameter $v$ is not a physical constant. The scheme is  very rigid
and it should be checked by the comparison with the experimental measures.
\par
The calculation has been performed in the Landau gauge
and by using the symmetric formalism whenever it was possible.
We checked the physical unitarity and the absence of $v$
in the measurable quantities.
\par
A very simple evaluation has been performed for the parameters of the classical
action, by using  leptonic processes. The parameter 
that describes the departure from the Weinberg relation between
$M_W$ and $M_Z$ is very small and the scale of the radiative corrections
is of the order of hundred GeV. That means that the model is on
solid grounds and it is reasonable to make further efforts for
the evaluation of the radiative corrections in other processes.


\section*{Acknowledgments}
One of us (R.F.) is honored to thank the warm hospitality of the
Center for Theoretical Physics at MIT, Massachusetts, where he
had the possibility to work partly on the present paper.


\appendix
\section{Limit $D=4$ for the logarithmic integral}
\label{app:A}

{We collect in this Appendix some relevant formulas}.
\begin{eqnarray}&&
\Delta_m \equiv \frac{1}{(2\pi)^D}\int d^D q\frac{i}{q^2-m^2}\,,
\nonumber\\&&
H(m,M)\equiv 
-\frac{1}{(2\pi)^D}\int d^D q\frac{1}{q^2-m^2}
\frac{1}{(p+q)^2-M^2}\,.
\label{se.2p}
\end{eqnarray} 
{The following identities allow to prove the cancellation of infrared 
     divergences due to the massless photon.} 
\begin{eqnarray}&&
G(M)\equiv \frac{\partial}{\partial m^2}H(m,M)\biggl|_{m^2=0}
\nonumber\\&& 
=\frac{i}{(4\pi)^\frac{D}{2}}\frac{\Gamma(3-\frac{D}{2})}{\Gamma(2)}
\int_0^1 dx(1-x)[M^2x-p^2x(1-x)]^{\frac{D}{2}-3}\, .
\label{va.17.1.2}
\end{eqnarray} 
\begin{eqnarray}&&
G(0)\equiv 
\lim_{M=0}G(M)
\nonumber\\&&  
=\frac{i}{(4\pi)^\frac{D}{2}}[-p^2]^{\frac{D}{2}-3}
\frac{\Gamma(3-\frac{D}{2})}{\Gamma(2)}
\frac{\Gamma(\frac{D}{2}-2)\Gamma(\frac{D}{2}-1)}{\Gamma(D-3)}\, .
\label{va.17.1.3}
\end{eqnarray} 
\begin{eqnarray}
\lim_{m=0}~\frac{1}{m^2}\Delta_m
=\lim_{m=0}~\frac{1}{(4\pi)^\frac{D}{2}}
\frac{\Gamma(1-\frac{D}{2})}{\Gamma(1)}
(m^2)^{\frac{D}{2}-2}=0 \,.
\nonumber\\&&  
\label{va.17.1.5}
\end{eqnarray} 
\begin{eqnarray}&&
\frac{\partial}{\partial M^2}G(M)\biggl|_{M=0}
\nonumber\\&&
=-\frac{i}{(4\pi)^\frac{D}{2}}\frac{\Gamma(4-\frac{D}{2})}{\Gamma(2)}
[-p^2]^{\frac{D}{2}-4}
\frac{\Gamma(\frac{D}{2}-2)\Gamma(\frac{D}{2}-2)}{\Gamma(D-4)} \,.
\label{va.17.1.8}
\end{eqnarray} 
%


For the integral in eq. (\ref{calc.13.1}) use the notation
\begin{eqnarray}
a=p^2,\quad  b= -p^2 +M^2-m^2,\quad c= m^2
\label{A.1}
\end{eqnarray} 
i.e.
\begin{eqnarray}
\int_0^1 dx 
\ln\biggl(p^2 x^2+[M^2-m^2-p^2] x +{m^2}\biggr)
=
\int_0^1 dx 
\ln\biggl(a x^2+b x +c\biggr).
\label{A.2}
\end{eqnarray} 
If one or two masses are zero one gets
\begin{eqnarray}&&
\int_0^1 dx 
\ln\biggl(x[p^2 x+M^2-p^2]\biggr)
\nonumber\\&&
=-2 +  \ln\biggl|p^2-M^2\biggr|
+\frac{M^2}{p^2}\ln\biggl|\frac{M^2}{p^2-M^2}\biggr|
-i\pi\frac{p^2-M^2}{p^2}\theta(p^2-M^2)
\nonumber\\&&
\label{calc.14.4}
\end{eqnarray} 
Let
\begin{eqnarray}
\Delta =[m^2+M^2-p^2]^2-4m^2M^2.
\label{A.3}
\end{eqnarray} 
By following the Feynman prescription one gets:
\par\noindent
for $0<p^2<(M-m)^2$, $ \Delta>0$ and then the integral is
\begin{eqnarray}
 -2 + \ln(a+b+c) + \frac{b}{2a} \ln\frac{(a+b+c)}{c}
+\frac{\sqrt{\Delta}}{2a}\ln\frac{2c+b-\sqrt{\Delta}}{2c+b+\sqrt{\Delta}};
\label{calc.21.2p}
\end{eqnarray}
\par\noindent
for $p^2>(M+m)^2$, $ \Delta>0$ and then the integral is
\begin{eqnarray}
 -2 + \ln(a+b+c) + \frac{b}{2a} \ln\frac{(a+b+c)}{c}
+\frac{\sqrt{\Delta}}{2a}\ln\frac{2c+b-\sqrt{\Delta}}{2c+b+\sqrt{\Delta}}
- i \frac{\sqrt{\Delta}}{a};
\label{calc.21.2p.x}
\end{eqnarray}

\par\noindent
for $p^2=(M-m)^2 $ or  $p^2=(M+m)^2 $, $\Delta=0$ and then the integral is
\begin{eqnarray}
 -2 + \ln(a+b+c) + \frac{b}{2a} \ln\frac{(a+b+c)}{c};
\label{calc.21.2p.4}
\end{eqnarray}
\par\noindent
for $(M-m)^2<p^2<(M+m)^2$, $ \Delta<0$ and then the integral is
\begin{eqnarray}&&
-2 -\frac{b}{2a}\ln c 
+(1+\frac{b}{2a})\ln[a+b+c]
\nonumber\\&&
+\frac{\sqrt{-\Delta}}{a}\Biggl\{\tan^{-1}
\biggl(\frac{2a+b}{\sqrt{-\Delta}}\biggr)-\tan^{-1}
\biggl(\frac{b}{\sqrt{-\Delta}}\biggr)
\Biggr\}
\label{calc.17p}
\end{eqnarray} 
where $-\frac{\pi}{2} <\tan^{-1}(x)<\frac{\pi}{2}, x\in{\cal R}$.



\begin{thebibliography}{99}


\bibitem{Bettinelli:2008ey}
  D.~Bettinelli, R.~Ferrari and A.~Quadri,
  ``The SU(2) X U(1) Electroweak Model based on the Nonlinearly Realized Gauge
  Group,''
  arXiv:0807.3882 [hep-ph], to appear in Int.\ J.\ Mod.\ Phys.\ A.

\bibitem{Bettinelli:2008qn}
  D.~Bettinelli, R.~Ferrari and A.~Quadri,
  ``The SU(2) X U(1) Electroweak Model based on the Nonlinearly Realized Gauge
  Group. II. Functional Equations and the Weak Power-Counting,''
  arXiv:0809.1994 [hep-th].

\bibitem{Bettinelli:2007tq}
  D.~Bettinelli, R.~Ferrari and A.~Quadri,
  Phys.\ Rev.\  D {\bf 77} (2008) 045021
  [arXiv:0705.2339 [hep-th]].

\bibitem{Bettinelli:2007cy}
  D.~Bettinelli, R.~Ferrari and A.~Quadri,
  Phys.\ Rev.\  D {\bf 77} (2008) 105012
  [arXiv:0709.0644 [hep-th]].

\bibitem{Bettinelli:2007eu}
  D.~Bettinelli, R.~Ferrari and A.~Quadri,
  J.\ Generalized Lie Theory and Applications
  {\bf 2} (2008), No 3, 122 [arXiv:0712.1410 [hep-th]].

\bibitem{Ferrari:2005ii}
  R.~Ferrari,
  JHEP {\bf 0508}, 048 (2005)
  [arXiv:hep-th/0504023].

\bibitem{Ferrari:2005va}
  R.~Ferrari and A.~Quadri,
  Int.\ J.\ Theor.\ Phys.\  {\bf 45}, 2497 (2006)
  [arXiv:hep-th/0506220].

\bibitem{Bettinelli:2007zn}
  D.~Bettinelli, R.~Ferrari and A.~Quadri,
  Int.\ J.\ Mod.\ Phys.\  A {\bf 23}, 211 (2008)
  [arXiv:hep-th/0701197].

\bibitem{Ferrari:2004pd}
  R.~Ferrari and A.~Quadri,
  JHEP {\bf 0411}, 019 (2004)
  [arXiv:hep-th/0408168].

\bibitem{Denner:1994xt}
  A.~Denner, G.~Weiglein and S.~Dittmaier,
  Nucl.\ Phys.\  B {\bf 440} (1995) 95
  [arXiv:hep-ph/9410338].

\bibitem{Grassi:1995wr}
  P.~A.~Grassi,
  Nucl.\ Phys.\  B {\bf 462} (1996) 524
  [arXiv:hep-th/9505101].

\bibitem{Grassi:1997mc}
  P.~A.~Grassi,
  Nucl.\ Phys.\  B {\bf 537} (1999) 527
  [arXiv:hep-th/9804013].

\bibitem{Grassi:1999nb}
  P.~A.~Grassi,
  Nucl.\ Phys.\  B {\bf 560} (1999) 499
  [arXiv:hep-th/9908188].

\bibitem{Ferrari:2000yp}
  R.~Ferrari, M.~Picariello and A.~Quadri,
  Annals Phys.\  {\bf 294} (2001) 165
  [arXiv:hep-th/0012090].

\bibitem{Becchi:1999ir}
  C.~Becchi and R.~Collina,
  Nucl.\ Phys.\  B {\bf 562} (1999) 412
  [arXiv:hep-th/9907092].


\bibitem{Amsler:2008zz}
  C.~Amsler {\it et al.}  [Particle Data Group],
  Phys.\ Lett.\  B {\bf 667}, 1 (2008).

\bibitem{Vilain:1994qy}
  P.~Vilain {\it et al.}  [CHARM-II Collaboration],
  Phys.\ Lett.\  B {\bf 335}, 246 (1994).


\bibitem{Bardin:1999ak}
  D.~Y.~Bardin and G.~Passarino,
  ``The standard model in the making: Precision study of the electroweak
  interactions,''
{\it  Oxford, UK: Clarendon (1999) 685 p}


\end{thebibliography}
\end{document}